# "NEW ZERO-RESISTANCE STATE" IN HETEROJUNCTIONS: A DYNAMICAL EFFECT

P W Anderson and W F Brinkman
Physics Department, Princeton University

Mani et al(1) and Zudov et al(2) have reported the observation of zero resistivity in heterojunctions at moderate magnetic fields(≈0.5 T) and low temperature (<1K), when the sample is irradiated with microwaves at a frequency somewhat higher than the cyclotron frequency $\Omega_C$ at the given field, and a large reduction in resistivity, or zero resistivity, at frequencies above the harmonics of $\Omega_C$. That there are oscillations in the longitudinal component of resistivity related to the cyclotron frequency has been reported earlier(3), but for the minima in these oscillations to be driven to zero requires low T and ultra-pure samples. Speculations as to the cause of this "zero-resistance state" have ranged widely; in (1), for instance, the suggestion is made of a form of superconductivity.

Our first point is that there is no demonstration that the state is one of zero resistivity; rather the observations suggest that the resistivity is oscillating PAST zero to negative values. This is entirely consistent with physical principles in a driven system. But a negative *local* resistivity is not easily observable, since it will lead to circulating currents which somehow dissipate the extra energy that is available, and will almost certainly be read macroscopically as zero resistance. There is some experimental evidence for circulating currents(R Willett, private comm.) Even so, a system with negative resistance is rare and understanding the current patterns and implied instability is interesting research. For example, negative resistance implies growth of charge density fluctuations that should be observable in the power spectrum of the current.

The second point is that the zero resistance appears as the end result of a continuous growth of the oscillations in resistivity, which except for the sudden cutoff at zero resistivity continue to increase in amplitude as the microwave power is further increased. The peaks in resistivity at $3/4\ \Omega_C$ and $1\ 3/4\ \Omega_C$ seem to be as large in the positive direction as the hypothetical negative continuation beyond zero at $5/4\Omega_C$ (see fig. 1) It is very hard to see how any type of superconductivity mechanism could increase the resistivity in this smooth, symmetrical way. In addition, it should be noted that the effect appears to be linear in the Microwave power at low powers and begins to saturate at higher powers.

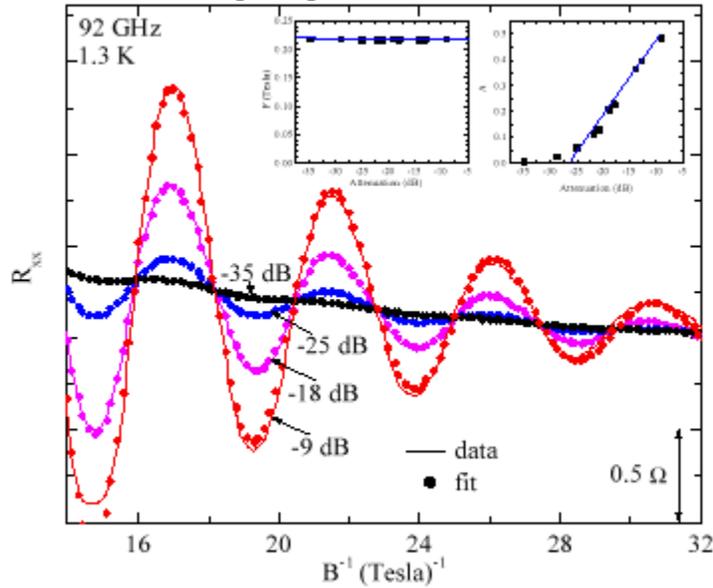

Figure 1 the reproduction of the data from reference 1. The longitudinal resistance is plotted versus the inverse magnetic field for different rf power levels.

A final notable point is that the resistivity at frequencies $n/2\, \Omega_C$ appears to remain almost exactly at the original value as the power is increased. One of the striking results in (1) and (2) is this crossing point of all curves at the cyclotron resonance and its harmonics, as well as the half-$\Omega_C$ points (see Fig 1, from ref 1). We should recall that in these samples, where the transport relaxation time is much longer than the cyclotron period, the Hall angle is nearly $\pi/2$: the current flows almost exactly along contour lines of the electric potential and the longitudinal resistivity rho$_{xx}$ is much smaller than the Hall resistivity $R_h \cdot B$. The observed phenomenon could be described as a rather small roughly periodic oscillation in the Hall angle, in the extreme cases passing beyond $\pi/2$ so that the current flows diagonally up the gradient of potential at the $5/4\, \Omega_C$ point, by roughly as much as it is driven down the gradient at $3/4\Omega_C$.

We propose the following speculative mechanism for this effect. It depends on the structure of the energy levels in crossed electric and magnetic fields. These can be written, in the gauge where the vector potential is A**x**=By,(y is also the direction of the electric field, so that the scalar potential is Ey) , in terms of the product of plane waves in the x direction, exp(ikx), times harmonic oscillator wave functions in the y direction centered about $y_O=k/B - E/B^2$. Their energies are given by

$$E_N(y_O) = (N+1/2)\Omega_C + Ey_O \qquad [1]$$

Thus at each $y_O$ there is a stack of Landau levels separated by $\Omega_C$ whose relative energies are displaced by the potential energy at that $y_O$. (see Fig 2) Current flows along the x direction predominantly, but because of scattering there is a slow net flow downhill, parallel to E, and the Fermi level tracks the electrostatic potential (and there is a little longitudinal conductivity.)

Now we hypothesize that the RF field can cause transitions among these levels, but that, of course, energy must be conserved. We realize that there are strong selection rules and that in fact all of the dipolar matrix elements are zero except between levels at the same $y_O$ separated by exactly $\Omega_C$. (This theorem is known as Kohn's theorem) But for the time being we ignore that fact and assume that all the levels can communicate with each other. Then we note that when $\Omega$ is slightly greater than $\Omega_C$, it connects an occupied level at $y_O$, with energy $(N+1/2)\Omega_C + Ey_O$, with a level at $(N+3/2)\Omega_C + E(y_O+\Delta y)$, *upfield* from the original wave function: while if the frequency is a little less than $\Omega_C$, the unoccupied level is d*ownfield*. (see Figure 2) Thus in the former case, we will tend to add negative resistivity, in the latter to add positive resistivity, since the only transitions we induce are to displaced wave-functions. There is no displacement at all at $\Omega_C$, and at $3/2\,\Omega_C$ the upfield and downfield rates will tend to be equal. Thus we seem to have acquired the basic spectroscopy of the problem, but so far there is a large hole in the argument:

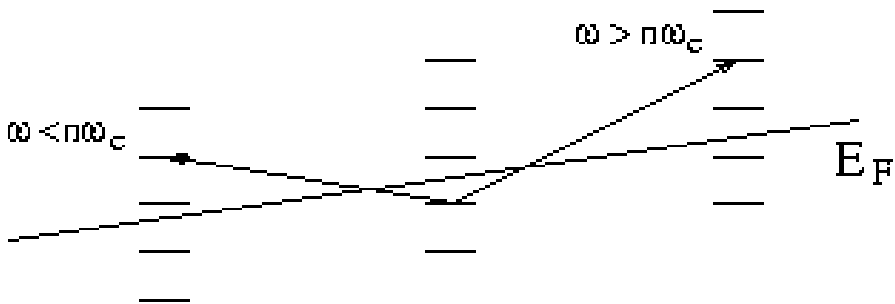

Figure 2 A schematic of the envisioned process for electron scattering up and down the potential field.

There are no matrix elements for these processes in the pure two-dimensional gas. The field involved in the argument must be the local field as the applied field is clearly not strong enough to make such transition probable. The result must therefore depend on the random potential that give rise to the residual resistivity. Attempts are being made to calculate this effect. In fact, a recent preprint from the Yale group has done a nice job of calculating the result.

The electron gas is known not to be pure in even the best heterojunctions. Even at high magnetic fields where the quantum Hall effect is observed, the landau levels are known to be broadened comparably to their width, so that the one-electron width is very large compared to the landau level separation at the low fields relevant to this phenomenon. On the other hand, the transport relaxation time, as evinced by the longitudinal conductivity or the widths of magnetoplasmon lines (Ong, 4), is quite long, the Q of these lines at low fields being >>1. The reason for this discrepancy is normally thought to be that the main component of the scattering is predominantly forward: there is a smooth random variation in the electrostatic potential V which distorts the electron orbits very considerably, in fact causing almost all of them to be localized around hills or valleys of V, only a single orbit at the center of the level being extended. The counterintuitive result is to decrease the real scattering between landau orbits and reduce $rho_{xx}$.

An equally reasonable point of view is that the *same* coherent backscattering can cause localization of almost all levels, even if the scattering is not mainly forward but is by random point scatterers, as demonstrated by Pruisken(5). In this case the actual orbits are random mixtures of a large number of the original Landau levels and we have exactly the situation which is called for to explain the observed microwave effect: there are effectively no selection rules, and any level may be connected to any other in its vicinity by a matrix element of the microwave field. It seems likely to us that the actual situation involves neither extreme case, but we suspect that so long as the variations of the potential are sufficiently rapid and random, the behavior will be the same.

How much of the level with a wave function belonging to quantum number N' will be mixed in to level N? We can approximate this using simple lowest-order perturbation theory, which gives us

$$G_N(\omega) = G_N^0(\omega) + G_N^0(\omega) \lambda^2 \sum G_N(\omega) \times G_{N'}(\omega)$$

Where the sum is taken over N' not equal to N and $G_N^0$ includes the intra-Landau level corrections due to scattering. The imaginary part of the second term coming from the N' Landau level is therefore

$$\text{Im}[G_N(\omega)] = \text{Re}[G_N(\omega)]^2 \times \text{Im}[G_{N'}(\omega)]$$

Since $\text{Re}[G_N(\omega)] \approx 1/(\omega - N\Omega_C)$ this gives us the matrix element decaying as a small negative power of $\omega - N\Omega_C$. This is consistent with the fall off of the oscillations in resistivity as one moves to higher harmonics.

These arguments indicate that the above is the physical nature of the observed oscillations of resistivity. It seems difficult to make a quantitative comparison without knowing more about the physical parameters such as the true microwave field level and the nature of the random potential. But, conversely, the phenomenon seems likely to tell us a great deal about the nature of conductivity in this regime which is as yet not well understood.

## Acknowledgement

The authors would like to acknowledge useful discussions with R. G. Mani, V Narayanamurti and R Willet as well as useful discussions with the Condensed Matter Theory group at Princeton.